# Using Page Size for Controlling Duplicate Query Results in Semantic Web

[1]Oumair Naseer, [2]Ayesha Naseer, [3]Atif Ali Khan, [4]Humza Naseer

[1,] School of Engineering, University of Warwick, Coventry, UK
[2,] Department of Computer Science, NUST, Islamabad, Pakistan
[3,] School of Engineering, University of Warwick, Coventry, UK
[4,] Business Intelligence Consultant, Bilytica Private Limited, Lahore, Pakistan

`[1]o.naseer@warwick.ac.uk, [2]ayesha.naseer@emc.edu.pk,`
`[3]atif.khan@warwick.ac.uk, [4]humza.naseer@bilytica.com`

**Abstract.** *Semantic web is a web of future. The Resource Description Framework (RDF) is a language to represent resources in the World Wide Web. When these resources are queried the problem of duplicate query results occurs. The present techniques used hash index comparison to remove duplicate query results. The major drawback of using the hash index to remove duplicate query results is that, if there is a slight change in formatting or word order, then hash index is changed and query results are no more considered as duplicate even though they have same contents. We presented an algorithm for detection and elimination of duplicate query results from semantic web using hash index and page size comparisons. Experimental results showed that the proposed technique removed duplicate query results from semantic web efficiently, solved the problems of using hash index for duplicate handling and could be embedded in existing SQL-Based query system for semantic web. Research could be carried out for certain flexibilities in existing SQL-Based query system of semantic web to accommodate other duplicate detection techniques as well.*

**Keywords***: Duplicate query results; Semantic web; Hash index; SQL-Based query system*

## 1. Introduction

The volume of data on web is growing day by day. World Wide Web has necessitated the users to locate their desired information resources and to assess their usage patterns [1]. The need for building server side and client side intelligent systems can mine for knowledge in a successful manner. Semantic web is the extension of the current web where semantics of information and services on the web are defined. Semantic web makes it possible for the web to understand and satisfy the requests of people and machines to use the web content [2]. The advancement of the semantic web has given rise to different problems related with it. One of the problems associated with semantic web is the duplicate query results. Multiple data sources referring to the same real entity leads to the problem of duplicate query results. A duplicate can be defined as; the exact same syntactic terms and sequences without formatting differences and layout changes. The current SQL-based approach to query Resource Description Framework RDF data on semantic web does not handle the problem of duplicate query results generated from multiple data sources [3].

In existing techniques [3, 4] duplicate query results are removed by computing and comparing the hash index of query results. The major drawback of using the hash index to remove duplicate query results is that, if there is a slight change in formatting or word order, then hash index is

International Journal of Web & Semantic Technology (IJWesT) Vol.4, No.2, April 2013

changed. Also the query results are no more considered as duplicate even though they have same contents.

This paper presented a technique to remove duplicate query results. It depends on distribution of document sizes and hash collisions will not be detected among documents of same sizes with slight change in formatting. The use of document size comparison along with hash value reduced the problems of using the hash index for duplicate detection. This technique has very low computational cost, reduced memory requirements of the system and can easily be embedded in existing SQL-Based query system of semantic web.

## 2. Related Work

In [5] author presented a novel technique for the estimation of the degree of similarity among pairs of documents. The approach was called shingling, which does not rely on any linguistic knowledge other than the ability to tokenize documents into a list of words, i.e., it is merely syntactic. In shingling, all word sequences of the adjacent words are extracted. If two documents contain the same set of shingles they are considered to be equivalent. If their sets of shingles appreciably overlap, then they are exceedingly similar. In order to reduce the set of shingles to a reasonably small size, authors employed an unbiased deterministic sampling technique that reduces the storage requirements for retaining information about each document. This also reduces the computational effort of comparing documents. The experiment was done on a set of 30 million web pages obtained from an AltaVista crawl. These pages were grouped into the clusters of extremely similar documents. A Compact features for the detection of near-duplicates in distributed retrieval is presented in [6]. In [7], author describes an approach to distinguish between different URLs having similar texts. For computer networks and ISDN systems [8] author describes an effective crawling through URL ordering. In [9], author describes an approach to find replicated web collections. Materialized view on oracle is presented in [10].

In [4] author proposed a pass-through architecture via hash techniques to remove duplicate query results. The system involved to receive the query from the user and issue the user query to a first data source. After receiving the result of first query from the first data source, the hash index is calculated for the first query result and result is passed to the user. Then the system further receives the results of second query and calculates the hash index of second query result. The first hash value is compared with the second hash value to check for the duplicate query results. If there is any hash collision then the first data source is queried to receive the results of second query, and if first data source contains data against second query then second query result is considered duplicated and discarded.

In [3], author proposed a SQL table function RDF_MATCH to query RDF data, which can search and infer from RDF data available on the semantic web. It also enables further processing by using standard SQL constructs [11-13]. The structure of the RDF_MATCH function [14-16] enables it to capture a graph pattern to be searched, RDF model and rule bases consisting of RDF data to be queried, and then provide the query results based on inference rules [17-19]. The RDF_MATCH function is implemented by generating SQL queries against tables that contain RDF data.

## 3. Methodology

Subject-property matrix materialized join views and indexes on data and rule bases are used to improve efficiency and further kernel enhancements have been provided to reduce the run time

50

overheads [20-24]. This query scheme efficiently retrieved RDF data on semantic web but did not handle the duplicate query results problem.

> **Step1: Issuing first query to first data source**
> In this very first step the system receives a first query from the user as input, after receiving the first query the system passes that query to the first data source to get the first query results in output.

> **Step2: Computation of the hash index and page size of first query result**
> In second step after receiving the results from first data source, we compute the hash index of the result. After computing the hash index we compute the page size of the first query result, then the calculated values of hash index and page size of first query result are stored in hash table along with the pointer to first data source. After storing this information in hash table the first query result is passed to the user.

> **Step3: Issuing second query to second data source**
> In third phase of this process the system receives a second query from the user as input, after receiving the second query the system pass that query to second data source to get the second query results in output.

> **Step4: Computation of the hash index and page size of second query result**
> In fourth step after receiving the results from second data source, we compute the hash index of the result. After computing the hash index we compute the page size of the second query result, then the calculated values of hash index and page size of second query result are stored in hash table along with the pointer to second data source.

> **Step5: Comparison of hash indexes**
> In this step the system proceeds by comparing the hash indexes of first and second query results. If the first hash index collides with the second hash index, then the first data source is queried for the results of the second query. If the first data source contains data in response to second query then second query results are considered as duplicate and are discarded.

> **Step6: Comparison of page sizes**
> In this phase we see, if the first and second hash indexes are not same then the page sizes of first and second query results are compared if the page sizes of first and second query results are same or vary within the threshold of 0 to 50 kb then the first data source is queried for the results of second query. If the first data source contains data in response to second query then second query results are considered as duplicate and are discarded.

## 3. Algorithm

The modified processing steps for the approach [3] can be written in an algorithmic form as under.

**ALGORITHM**: Building a module to remove duplicate query results from semantic web.
**INPUT**: Query Results of RDF data from a semantic data source.
**OUTPUT**: Query Results free of duplicates
**STEP1**: /* Displaying the First Query Results to User After Receiving the Query.
**DO**   Read the user query/ problem statement.

    1.2:  WRITE Query to Data Source.
    1.3:  READ Query Results
    1.4:  COMPUTE Hash Index
    1.5:  COMPUTE Page Size
    1.6:  SAVE the Hash Index, Page Size and Pointer In Hash Table.
    1.7:  DISPLAY results to the user.
**STEP2**: /* Displaying the Further Query Results to User after Removing duplicate results.

    2.1: **WHILE not end of** query results   **DO**
    2.2:     READ the Query
    2.3:  COMPUTE Hash Index
    2.4:  COMPUTE Page Size
    2.5**: IF**   Hash Index values in STEP 2.3 IS EQUAL to the Hash Index in STEP 1.4 DISCARD the result.
    2.6**:  ELSE IF** page size values in STEP 2.4 IS EQUAL to the page size in STEP 1.5 DISCARD the result.

    **ELSE    GOTO** STEP 1.6
           {END IF}
 END {DO WHILE}

## 3. Query Handling

The detailed description for the process is as under:

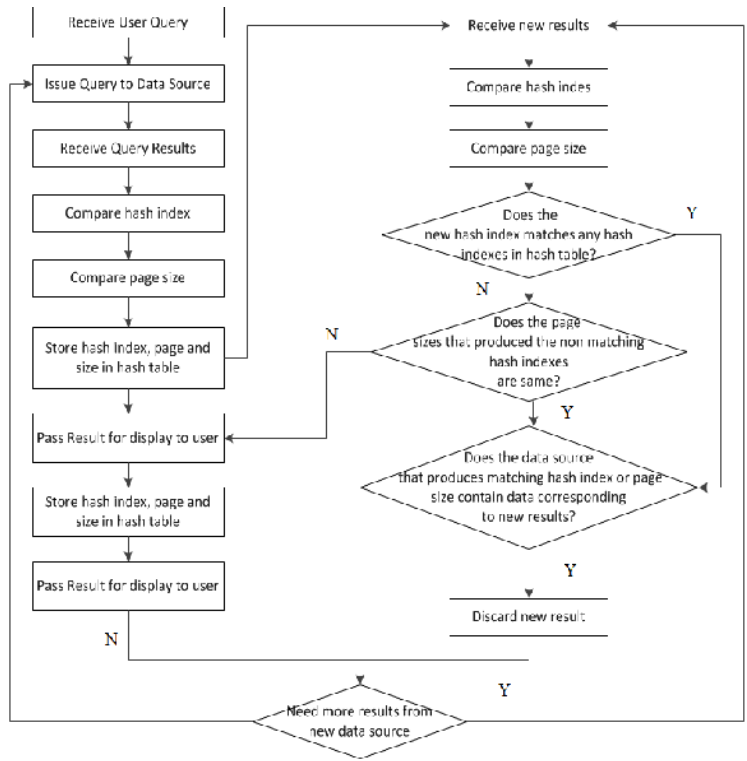

Figure 1: Dataflow diagram of proposed algorithm.

The proposed algorithm involves receiving the first query from the user and issuing the user query to a first data source. After receiving the result of first query from the first data source the hash index and page size is calculated for the first query result and result is passed to the user. Then the system further receives the results of second query and calculates the hash index and page size of second query result. The first hash value is compared with the second hash value to check for duplicate query results. If there is any hash collision then the first data source is queried to receive the results of second query, and if first data source contains data against second query then second query result is considered duplicate and is discarded. If the first and second hash indexes are not same then the first page size is compared with second page size. If the page sizes are same or vary within the threshold of 0 to 50 kb then again the first data source is queried to receive the results of second query. And if first data source contains data against second query then second query result is considered as duplicate and is discarded. The proposed technique is an enhancement of SQL-Based scheme to query RDF data on semantic web; it extends its functionality to remove duplicate query results.

## 4. Results

Now we shall give the results of our proposed technique. The success of the proposed technique depends on the distribution of the document sizes and hash collisions will not be detected among documents of same size with slight changes in formatting and word order. We performed an experiment on data set of 16 web pages collected randomly; the links of collected web pages are given in following tables. We calculated the values of hash index and page sizes of data set. Then we run our proposed algorithm on this data set which updates formatting and word order of web pages. Again, we computed the values of hash index and page sizes of same data set. Following tables (1, 2) and graphs (1, 2) showed the results of computed values and their variations.

**Table 1:** Download links of web pages, their corresponding hash values and page sizes.

| Serial No. | Download Links of Web Pages | MD5 Hash Values | Page Size |
|---|---|---|---|
| 1 | http://download.cnet.com/WinRAR-32-bit/3000-2250_4-10007677.html | 123D463728394FDB876A6534B | 31.1KB |
| 2 | http://www.hotmail.co.uk | 35645623FBD8786A667F76B8A | 20.5KB |
| 3 | https://mail.google.com/mail/u/0/?shva=1 | 4565778898965320B98A9F5D12 | 20.7KB |
| 4 | http://www.winzip.com/index.htm?sc_cid=go_uk_b_search_wz_brand | 768735D6544A4433B767F24432 | 61.5KB |
| 5 | http://download.cnet.com/Skype/3000-2349_4-10225260.html | 87B68634AF8768B76767D87676 | 265KB |
| 6 | http://www.google.com | 40877123B656D565F564356A11 | 368KB |
| 7 | http://vlc-media-player.todownload.com/?gclid=CO7D44CR0bQCFaTKtAodbg0ATA | 3635B78987F56567D434A989B1 | 87.5KB |
| 8 | http://www.youtube.com | 837735299BFD26536723A565FA | 5.57KB |
| 9 | http://www.uet.edu.pk | 9897860B798790D098A454F767 | 8.57KB |
| 10 | http://www.mcs.edu.pk | 365236B6767F5465A766D5767F | 3.66KB |
| 11 | http://www.rect.edu.pk | 213625371825637BFDA768C76C | 24.5KB |
| 12 | http://www.jackson.com | 5656C67B8768F97898D7878A19 | 32.7KB |
| 13 | http://www.w3.org/TR/rdf-primer/ | 986765320BCFA565DF6767C767 | 65.4KB |
| 14 | http://www.bru.com | 386827368BFDAC79878CDA897 | 92.3KB |
| 15 | http://www.education.com | 213567153BFDA987897CFCA92 | 89.7KB |
| 16 | http://www.baqai.com | 387867FCDA87878CFDA7887B6 | 91.9KB |

**Table 2:** Hash values and page sizes of web pages after changes in formatting and word order.

| Serial No: | Duplicate Web Page | MD5 Hash Values | Page Sizes |
|---|---|---|---|
| 1 | WinRAR Web Installation Page | 27BA8126354FCD345235CF | 31.1KB |
| 2 | Hotmail Web Page | 123546AAF786534Act67325 | 20.5KB |
| 3 | Gmail Web Page | 2443547FCA544335DF678A | 20.7KB |
| 4 | WinZip Web Installation Page | 10293645ACF534627DCB878 | 61.5KB |
| 5 | Skype Web Page | 23454789CFD4234ACB878F | 265KB |
| 6 | Google Web Page | 34546678FC5D4A099B2431 | 368KB |
| 7 | VLC Web Installation Page | 12343241CD6789AD98976B | 87.5KB |
| 8 | YouTube Web Page | 3546728CFB8767B6754FA6 | 5.57KB |
| 9 | UET Web page | 234DCF655F665A776B556A | 8.57KB |
| 10 | MCS Web Page | 778644AA5F5DD4434B56AF | 3.66KB |
| 11 | RECT Web Page | 2341325364786FCDA565DF5 | 24.5KB |
| 12 | Jackson Web Page | 0902365542536ADF675B78A | 32.7KB |
| 13 | RDF Premier Page | 534098FB7653455AD89B7D | 65.4KB |
| 14 | Bru Web Page | 23416354688439DCF67A5BF | 92.3KB |
| 15 | Education Web Page | 2234B564840986AF37846BA | 89.7KB |
| 16 | Baqai Web Page | 455463787846546BFCAD677 | 91.9KB |

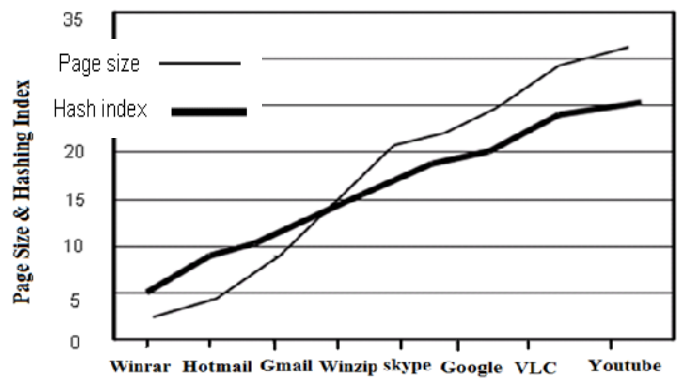

Figure 2: Variation of page size and hash values of collected data set of original web pages before formatting changes.

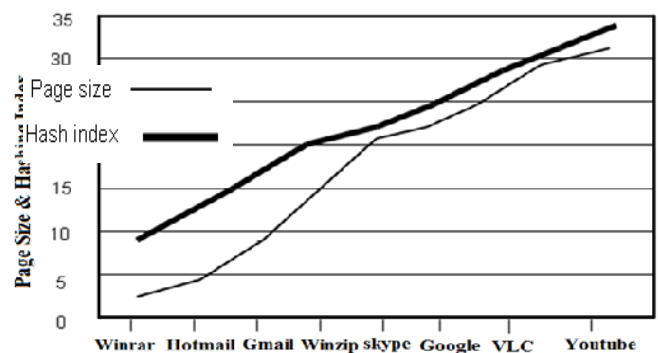

Figure 3: Variation of page size and hash values of collected data set of web pages with formatting changes.

In graph 1 the thick line represents the hash indexes of original web pages, and thin line represents the page size values of original web pages. In graph 2 the thick line represents the hash values of collected web pages after formatting changes, while the thin line shows the values of page sizes of web pages after formatting changes. By the comparison of both graphs it is concluded that the page size values of web pages remain the same after formatting changes as shown by thin line in graph 1 and graph 2, while the hash indexes are changed with formatting changes in the web pages as shown by the thick lines in graph 1 and graph 2. So the above graphs show that with formatting changes of web pages the page sizes remain similar while the values of hash indexes changed even though the contents of web pages are similar. It is concluded that using the page size comparison along with hash index solve the problems of just using the hash index for duplicate detection. It does not make the system computationally complex, reduces the memory requirements of the system and can easily be embedded in existing SQL-Based query system of semantic web.

## 5. Conclusions

A technique to remove the duplicate query results on semantic web using hash index and page size comparison has been presented in this paper. The proposed technique involves receiving the query from the user and issuing the user query to a first data source. After receiving the result of first query from the first data source the hash index and page size are calculated for the first query result and result is passed to the user. Then the system further receives the results of second query and calculates the hash index and page size of second query result. The first hash value is compared with the second hash value to check for the duplicate query results. If there is any hash collision then the first data source is queried to receive the results of second query, and if first data source contains data against second query then second query result is considered as duplicate and is discarded. If the first and second hash indexes are not same then the first page size is compared with second page size. If the page sizes are same or vary within the threshold of 0 to 50 kb then again the first data source is queried to receive the results of second query, and if first data source contains data against second query then second query result is considered duplicate and is discarded.

## 6. Future Recommendations

It is concluded that using the page size comparison along with hash index solve the problems of just using the hash index for duplicate detection. It does not make the system computationally complex, reduces the memory requirements of system. The proposed technique is an enhancement of SQL-Based scheme to query RDF data on semantic web; it further extends its functionality to remove duplicate query results.

Research can be carried out for certain flexibilities in existing SQL-Based query system of semantic web to accommodate other duplicate detection techniques as well. The concept of optimizing self-join queries that usually occur when querying RDF data can improve the efficiency of query processing in semantic web.

[5] Broder, A., Glassman, S., Manasse, M., and Zweig, G. "Syntactic clustering of the Web", In 6th International World Wide Web Conference, pp: 393-404, 1997

[6] Bernstein, Y., Shokouhi, M., and Zobel, J. "Compact Features for Detection of Near- Duplicates in Distributed Retrieval", in 'Proceedings of String Processing and Information Retrieval Symposium (to appear)', Glasgow, Scotland, 2006.

[7] [BarYossef, Z., Keidar, I., Schonfeld, U.. "Do Not Crawl in the DUST: Different URLs with Similar Text", 16th International world Wide Web conference, Alberta, Canada, Data Mining Track, 8-12 May, 2007.

[8] Cho, J., Garca-Molina, H., and Page, L. "Efficient crawling through URL ordering", Computer Networks and ISDN Systems, Vol. 30, No. 1-7, pp: 161-172, 1998.

[9] Cho, J., Shivakumar, N., Garcia-Molina, H. "Finding replicated web collections", ACM SIGMOD Record, Vol. 29, No. 2, pp. 355 - 366, June, 2000.

[10] Bello, Randall G., et al. "Materialized views in Oracle." PROCEEDINGS OF THE INTERNATIONAL CONFERENCE ON VERY LARGE DATA BASES, Vol. 24, INSTITUTE OF ELECTRICAL & ELECTRONICS ENGINEERS, 1998.

[11] Taylor, Chris. "An introduction to metadata." University of Queensland Library: 07-29, 2003.

[12] Bizer, Christian, and Richard Cyganiak. "D2r server-publishing relational databases on the semantic web." 5th international Semantic Web conference, 2006.

[13] Karvounarakis, Gregory, et al. "RQL: a declarative query language for RDF." Proceedings of the 11th international conference on World Wide Web, ACM, 2002.

[14] Wilkinson, Kevin, et al. "Efficient RDF storage and retrieval in Jena2." Proceedings of SWDB. Vol. 3, 2003.

[15] Haase, Peter, et al. "A comparison of RDF query languages." The Semantic Web–ISWC2004 502-517., 2004.

[16] Miller, Libby, Andy Seaborne, and Alberto Reggiori. "Three implementations of SquishQL, a simple RDF query language." The Semantic Web—ISWC 2002 (2002): 423-435., 2002.

[17] Prud, Eric, and Andy Seaborne. "Sparql query language for rdf." ,2006.

[18] Shivakumar, Narayanan, and Hector Garcia-Molina. "Finding near-replicas of documents on the web." The World Wide Web and Databases: 204-212.1999.

[19] Shivakumar, Narayanan, and Hector Garcia-Molina. "SCAM: A copy detection mechanism for digital documents."1995.

[20] Denehy, Timothy E., and Windsor W. Hsu. "Duplicate management for reference data." Computer Sciences Department, University of Wisconsin and IBM Research Division, Almaden Research Center, Research Report 2003.

[21] Honicky, R. J., and Ethan L. Miller. "A fast algorithm for online placement and reorganization of replicated data." Parallel and Distributed Processing Symposium, 2003. Proceedings International IEEE, 2003.

[22] Gomes, Daniel, André L. Santos, and Mário J. Silva. "Managing duplicates in a web archive." Proceedings of the 2006 ACM symposium on applied computing. ACM, 2006.

[23] Narayana, V. A., P. Premchand, and A. Govardhan. "Effective Detection of Near Duplicate Web Documents in Web Crawling." International Journal of Computational Intelligence Research 5.1,: 83-96., 2009.

[24] Su, Weifeng, Jiying Wang, and Frederick H. Lochovsky. "Record matching over query results from multiple web databases." Knowledge and Data Engineering, IEEE Transactions on 22.4: 578-589, 2010.
Corresponding Author:
Oumair Naseer
Department of Computer Science,
University of Warwick, Coventry, United Kingdom
o.naseer@warwick.ac.uk
56